\documentstyle[preprint,aps,epsf]{revtex}       
\global\let\epsfloaded=Y 
\begin{document}
\pagestyle{empty}                                
\preprint{
\font\fortssbx=cmssbx10 scaled \magstep2
\hbox to \hsize{
\hfill $
\vtop{
 \hbox{ }}$
}
}
\draft
\vfill
\title{
Higgs Mass from Casimir Energy Induced 
Cosmological Constant in the Standard Model
}
\vfill
\author{
Xiao-Gang He}
\address{
Department of Physics, National Taiwan University,
Taipei, Taiwan}
%
%
\vfill
\maketitle
\begin{abstract}
Casimir vacuum energy is divergent. It needs to be regularized. 
The regularization introduces a 
renormalization scale which may lead to a scale dependent 
cosmological constant. 
We show that the requirement of physical cosmological constant
is renormalization scale independent provides important 
constraints on possible 
particle contents and their masses in particle physics models. 
In the Standard Model of strong and 
electroweak interactions, besides the Casimir vacuum energy there is also 
vacuum energy induced 
from spontaneous symmetry breaking. 
The requirement that the total vacuum energy to be scale independent 
dictates the Higgs mass to be
$m_H^4 = 8\sum_i N_i m^4_i - 12m^4_W - 6 m^4_Z$, where the summation is
over fermions and $N_i$ equals to 3 and 1 for quarks and leptons, respectively.
The Higgs mass is predicted to be approximately $382$ GeV. 
\end{abstract}
%
%
\pacs{PACS numbers:  
 }
%
%
\pagestyle{plain}

Recent data from Type Ia supernovae  
indicate that our universe is in an accelerating expansion phase\cite{1}. 
Accelerating expansion implies the existence of 
energy form(s) (dark energy) 
with equation of state parameter $\omega = 
(pressure\;density/energy\;density)$ 
to be less than $-1/3$. The cosmological 
constant $\Lambda$, with $\omega = -1$, is one of the 
leading candidates for the dark energy. 
Combining information from Cosmic Microwave Background radiation, 
cluster survey and other data, 
the energy density
provided by the cosmological constant relative to 
the critical energy density
is determined to be approximately $65\%$\cite{2}. 
Implications for a universe with
cosmological constant have been extensively studied\cite{2a,3}.
The cosmological constant $\Lambda$ is related to the physical
vacuum energy density $\rho_p$ by $\Lambda = 8 \pi G_N \rho_p$. 

At present there are no convincing theoretical models
which can produce the observed size of the cosmological constant. 
This is one of the most prominent problems in cosmology and particle 
physics\cite{3}.
There are
several discussions about constraints on properties of the 
Standard Model (SM) of the 
strong and electroweak interactions from vacuum energy. 
Most of them tried to estimate the size of the cosmological constant and
how it changes with energies\cite{4,5,6}.  
There are also several quantum field theoretical studies of the vacuum energy
renormalization properties\cite{od}.
In this paper we will not attempt to solve the cosmological problem, 
except to say that whatever mechanism producing the physical 
cosmological constant, it must have a value which agrees with observation. 
Instead we will consider possible constraints
on particle physics model parameters, such as particle contents and 
particle masses using the fact that the physical cosmological 
constant, being a physical parameter, must be 
renormalization scale independent. We show that this simple requirement 
can provide important insight about a long standing problem for particle 
physics, the Higgs mass in the Standard Model.

In quantum field theory, there are loop corrections to the vacuum energy which
are some times divergent such as the contribution from the 
Casimir vacuum energy 
(vacuum energy due to particle fluctuation in the background), 
and the induced vacuum energy from spontaneous symmetry breaking. 
The divergent part of the vacuum energy must be renormalized to have physical 
meaning for quantum corrections. Renormalization in general 
introduces 
renormalization scale dependent vacuum energy. The cosmological 
constant being physical observable must not
depend on the renormalization scale. 

The Standard Model 
based on the gauge group $SU(3)_C\times SU(2)_L\times U(1)_Y$ is a 
very successful theory. Many of
its properties have been studied and verified experimentally, 
but one of the most important particle of the model, 
the Higgs particle, has yet to be 
discovered. Present experimental lower bound on the Higgs 
mass is $114$ GeV\cite{7}. Electroweak precision data prefers a 
light Higgs mass around the present lower limit\cite{8}. 
But larger mass up to several hundred GeV is allowed. 
There are also some bounds from various theoretical considerations\cite{9}. 
It is important if the Higgs mass can be constrained or
determined from some processes.
We find that the Higgs mass in the 
SM can indeed be constrained from 
the requirement that the total vacuum energy from Casimir
vacuum energy and induced vacuum energy from spontaneous symmetry breaking 
to be independent of the renormalization scale. We find that in the SM 
the Higgs mass must be close to $382$ GeV. Higgs mass as low as $382$ GeV
can be studied at the CERN Large Hadron Collider (LHC). The prediction
will be tested.  

We now provide more details on how information on Higgs mass can be 
obtained from the study of vacuum energy in the Standard Model.
We will first discuss the renormalization scale dependence of Casimir
vacuum energy and then discuss the scale dependence of vacuum energy in
the SM, and the implications for the Higgs boson mass.
The Casimir vacuum energy density 
produced by one degree of freedom of a massive 
boson particle is given by

\begin{eqnarray}
\rho_C = {1\over 2} \int {d^3k \over (2\pi)^3} \sqrt{k^2 + m^2}.
\label{ce}
\end{eqnarray}

It is clear that $\rho_C$ is divergent which 
must be regularized. To regularize the
divergent part in this integral, 
we use a dimensional regularization by changing the
integral to $n = 3-2\epsilon$ with $\epsilon$ to be taken to zero at the 
end\cite{6}

\begin{eqnarray}
\rho_C ={\mu_0^{2\epsilon}\over 4\pi^2} \int k^{1-2\epsilon} d k^2 
\sqrt{k^2 + m^2}.
\end{eqnarray}
Here the $\mu_0$ parameter is introduced to make sure that the dimension of the
integral is the same as in eq.(\ref{ce}). We have

\begin{eqnarray}
\rho_C = -{m^4\over 64 \pi^2}({1\over \epsilon} + 2 \ln 2 -{1\over 2} 
+ \ln {\mu^2\over m^2}) + {m^4\over 32 \pi^2} \ln{\mu^2\over \mu^2_0}.
\end{eqnarray}
The first term in the above is to be renormalized away. 
The second term is $\mu_0$ dependent. One should also note 
that in the above prescription the vacuum energy density 
of a zero mass particle has been normalized to be zero.

From this procedure it is not possible to fix the value of the total 
vacuum energy produced because the bare vacuum energy density is 
unknown, 
any value can be obtained. As have been mentioned that in this paper 
we will not attempt to determined
the value of the net cosmological constant, but 
to study possible constraints
on particle physics model parameters by requiring that the physical 
vacuum energy must be $\mu_0$ independent.

For a given theory there are more than one particle degrees of freedom. 
The total Casimir vacuum energy density is the sum of 
contributions from all particles. When 
summing over all particle contributions, one should multiply a ``minus'' sign
for the contribution from a fermion particle degree of freedom because the
anti-commuting nature of fermions. One has

\begin{eqnarray}
\rho_C^{eff} ={1\over 32 \pi^2} \ln{\mu_0^2\over \mu^2} {1\over 2}( 
\sum_{i=fermions} m^4_i - \sum_{i=bosons}m^4_i).
\end{eqnarray}
In the above we have removed the part which is not $\mu_0$ dependent and
have referred to the remaining vacuum energy as $\rho_C^{eff}$.
It is clear that if a theory only has boson or fermion degree of freedom, 
it is not
possible to have a scale $\mu_0$ independent $\rho_C^{eff}$. 
However when appropriate fermion and boson degrees of freedom 
appear in the theory, cancellations can happen such that the net 
cosmological constant produced by the Casimir vacuum energy is scale
$\mu_0$ independent. 

If the universe is supersymmetric, the above effective energy density
$\rho_C^{eff}$ 
is automatically zero, therefore the total cosmological constant, is
scale independent because in supersymmetric theories there are 
equal numbers of
fermion and boson degrees of freedom, and each bosonic particle and its 
fermionic supersymmetric 
partner have the same mass.
In fact this not only makes the cosmological constant 
scale independent, it also
dictates the cosmological constant to be exactly zero. However, 
supersymmetry is known to be broken, scale dependence will be introduced by
supersymmetry breaking effect in general unless there are certain 
relations between   
mass spectrum in a given model to facilitate the cancellation.

Without supersymmetry, elimination of  the scale dependence of the cosmological constant is
also possible if there are relations between particle masses and have the
right number of degrees of freedom.
In the SM, the particle contents are fixed. Cancellation of the scale dependence
can only be caused by relation between the particle masses. In the 
SM, $\rho_C^{eff}$ is given by
\begin{eqnarray}
\rho_C^{eff} = {1\over 32 \pi^2} \ln{\mu^2\over \mu_0^2}
{1\over 2}(m^4_H + 6 m^4_W + 3 m_Z^4 - 4\sum_i N_i m^4_i),
\label{casimir}
\end{eqnarray}
where the summation is over the SM fermions with $N_i$ being 3 and 1 for quarks and charged leptons, respectively. For neutrinos depending on whether they
are massive Dirac particles or Majorana particles, $N_i = 1$ or $N_i =1/2$.
If the above is the only contribution to the total vacuum energy, 
the elimination of the 
dependence on $\mu$ requires the existence of a mass
relation between the Higgs boson mass and the other known particle masses.
We obtain

\begin{eqnarray}
m^4_H = 12 m^4_t - 6 m^4_W - 3 m_Z^4 + O(m^4_b, ...),
\label{hm}
\end{eqnarray}
It is interesting to see that the above provides a way to determine the 
Higgs mass which is otherwise unknown in the SM. Using the known 
values\cite{10} 
$m_Z = 91.2$ GeV, $m_W = 80.4$ GeV, and $m_t = 174.3$ GeV, the Higgs
particle should have a mass about $321$ GeV.

One can apply the formula to the Two Higgs doublet model. In that case there
are five physical Higgs bosons. One would obtain

\begin{eqnarray}
\sum_{i=1,2,3,4,5} m^4_{H_i} = 12 m^4_t - 6 m^4_W - 3 m_Z^4 + O(m^4_b, ...),
\end{eqnarray}
From the above one finds that at least one of the Higgs boson must have a mass
less than $215$ GeV.  

The effective Casimir vacuum energy obtained in eq. (\ref{ce}) 
can be generalized
to models beyond the SM with more particles and with different gauge groups. 
However the
predictive power for the Higgs mass will be less if the new particle masses
introduced are not known. 

The above predictions are based on the assumption that the Casimir vacuum 
energy is the only source for vacuum energy. However in the SM there are 
additional source contributing to the vacuum energy through spontaneous symmetry 
breaking due to Higgs field condensation. 
The Higgs potential in the SM is given by

\begin{eqnarray}
V(H) = V_{vac-bare} + \mu^2 H^\dagger H + {1\over 2}
\lambda (H^\dagger H)^2.
\end{eqnarray}
Here $H^\dagger = (v + H^0, \sqrt{2}H^+)/\sqrt{2}$ 
is the Higgs doublet in the SM. The non-zero 
vacuum expectation value $v$ breaks 
the $SU(2)_L\times U(1)_Y$ to $U(1)_{em}$.
The term $V_{vac-bare}$ is a constant (no scale dependence) 
representing the bare vacuum energy 
which can be used to renormalize any divergent vacuum energy generated at loop
level. After spontaneous electroweak symmetry breaking, one obtains a 
vacuum energy given by

\begin{eqnarray}
\langle V\rangle = V_{vac-bare} - {m^4_H\over 8 \lambda}.
\end{eqnarray}
The imaginary part of $H^0$ and $H^+$ are ``eaten'' by the Z and W 
bosons, and the real part of $H^0$ becomes massive with a mass $m_H$
given by $m^2_H = \lambda v^2$. 
$\langle V\rangle $ is the classic physical vacuum energy.

When quantum corrections are included, 
there are additional contributions at loop level. 
There are two types of quantum corrections. 
One is the Casimir vacuum energy
discussed earlier. This vacuum energy can be viewed as quantum fluctuation
with closed boson and fermion loops in the vacuum which adds to 
$\langle V \rangle$ an additional term $V_{loop}$. 
The other is quantum corrections to the Higgs mass $m_H$ and the coupling
$\lambda$. These corrections are divergent and need regularization. 
This leads to 
a renormalization $\mu$ scale dependent cosmological constant. One 
needs to find
the conditions that can make the physical vacuum energy to 
be renormalization scale $\mu$ independent. 
To this end we have evaluated the $\beta$ function of the vacuum energy. 
We have

\begin{eqnarray}
(4\pi)^2 \mu {\partial \langle V \rangle \over \partial \mu} =
(4\pi)^2 \mu {\partial \over \partial \mu}(V_{loop} - {m^4_H\over 8 \lambda}) 
={1\over 2} m^4_H + 6 m^2_W + 3 m^4_Z - 4 \sum_i N_im^4_i.
\label{ph}
\end{eqnarray}
In obtaining the above equation we have used\cite{4}

\begin{eqnarray}
&&(4\pi)^2 \mu {\partial V_{loop} \over \partial \mu} = {1\over 2} m^4_H
+ 3 m^4_W + {3\over 2} m^4_Z - 2 \sum_i N_i m^4_i,\nonumber\\
&&(4\pi)^2 \mu {\partial\over \partial \mu} (-{m^4_H \over 8\lambda})
=(4\pi)^2 \mu ( - {m^2_H\over 4 \lambda} {\partial m^2_H\over \partial \mu} + 
{m^4_H\over 8 \lambda^2} {\partial \lambda \over \partial \mu}),
\end{eqnarray}
with

\begin{eqnarray}
&&(4\pi)^2 \mu {\partial m^2_H\over \partial \mu} = 
m^2_H (6 \lambda - {9\over 2} g^2 - {3\over 2} g'^2 + 2 \sum_i N_i h^2_i)
\nonumber\\
&&(4\pi)^2 \mu {\partial \lambda \over \partial \mu} = 12 \lambda^2 - 9\lambda  g^2
- 3 \lambda g'^2 + {9\over 4} g^4 + {3\over 2} g^2 g'^2 + 
{3\over 4} g'^4 + 4 \sum_i N_i h^2_i(\lambda - h^2_i),
\end{eqnarray}
where $g$, $g'$ and $h_i$ are the $SU(2)_L$, 
$U(1)_Y$ and the Yukawa couplings, respectively. 
The particle masses are given by $m_W^2 = g^2v^2/4$, $m^2_Z = (g^2+g'^2)v^2/4$
and fermion masses $m_i = h_i v/\sqrt{2}$. 

If the right hand side of eq. (\ref{ph}) does not vanish, the total 
vacuum energy density
$\langle V\rangle$ 
depends on the scale $\mu$. In order to have a $\langle V \rangle$ 
independent
of the scale, one must demand

\begin{eqnarray}
m^4_H = 8\sum_i N_i m^4_i - 12 m^4_W - 6 m^4_Z
\end{eqnarray}
One notes that the right hand side of 
the above equation is twice that of eq. (\ref{hm}). 
Therefore the Higgs mass determined from the above equation is
larger by a factor of $2^{1/4}$. The use of eq. (\ref{hm}) 
is not complete. We predict that the Higgs mass in the 
SM must be around

\begin{eqnarray}
m_H \approx 382 \mbox{ GeV}.
\end{eqnarray}

The Higgs mass determined this way is more than a factor of three 
larger than the
present lower bound, but is within the reach of LHC. The prediction 
can be tested in the near future.

The above result is obtained at the one loop level. There are 
higher order loop 
corrections. These corrections will shift the predicted value for
the Higgs mass, but the change will be small. The Higgs mass is therefore
expected to be around $382$ GeV.

There is also the possibility that the bare vacuum energy $V_{vac-bare}$
used to renormalize the divergent part also depends on the scale 
such that this scale dependence cancels exactly the one due to loop
corrections discussed in the above\cite{od}. 
In that case there is no information
about Higgs mass can be obtained. At present, we do not see how this can
occur in the Standard Model by directly evaluating the known contributions 
to the beta function. The approach taking here is to force the known
beta function to be zero without adding new terms in the Lagrangian. Further
study to understand the vacuum energy 
renormalization properties are necessary. At the phenomenological level, the
approach taking here is, in some way, to require that the theory  should 
not have quartic divergences which is similar in spirit to some of the 
studies in Ref.\cite{9} to constraint particle masses by requiring certain
divergences to vanish.   

One can apply the method to extensions of the SM. 
The generalization of the Casimir vacuum energy part is trivial. One just
needs to specify the particle contents and their masses. The scale 
dependence of the induced vacuum
energy part, terms of the form $-m^4_H/8 \lambda$, is however model dependent.
One needs more information about the model parameters to obtain detailed 
numerical numbers for the Higgs mass or any other parameters. Nevertheless,
the requirement of the total vacuum energy to be scale independent still 
provides an important constraint on models.

In conclusion, we have shown that constraints on particle physics model
parameter can be obtained from
the requirement of the physical vacuum energy to be renormalization scale
independent. In the Standard Model this simple requirement predicts that the
Higgs mass is approximately 382 GeV. 
This prediction can be tested in the near 
future at LHC. Confirmation of this prediction will provide important 
information about the Higgs mechanism in the Standard Model for strong and
electroweak interactions, it may also provide an important clue to solve
the cosmological constant problem.

{\bf\large Acknowledgments}

This work
was supported in part by 
National Science Council under grants NSC
91-2112-M-002-42,
and in part by the Ministry of
Education Academic Excellence Project 89-N-FA01-1-4-3.

\end{document}